\documentstyle[12pt, aaspp4]{article}
\begin{document}


\title{New Photometry for the Intermediate-age LMC Globular Cluster NGC 2121 and
the Nature of the LMC Age Gap$^1$}

\author{R. Michael Rich\altaffilmark{2}, Michael M. Shara\altaffilmark{3}, and David Zurek\altaffilmark {3} }

\altaffiltext{1}{Based on observations with the NASA/ESA {\it Hubble
Space Telescope}, obtained at the Space Telescope Science Institute,
which is operated by the Association of Universities for Research in Astronomy,
Inc, under NASA contract NAS 5-26555.}

\altaffiltext{2}{Department of Physics and Astronomy, UCLA, 8979 Math-Sciences Building, Los Angeles,
California  90095-1562; rmr@astro.ucla.edu}
\altaffiltext{3}{Department of Astrophysics, American Museum of Natural History, Central Park West
and 79th St.  New York, NY  10024-5192; mshara@amnh.org}

\begin{abstract}

We report new photometry for the cluster NGC 2121
in the Large Magellanic Cloud, which shows
a prominent hydrogen core exhaustion gap at the turnoff, and a
descending subgiant branch reminiscent of Galactic open clusters.
We achieve
an excellent fit using the Girardi isochrones, finding
an age of $3.2\pm 0.5$ Gyr, with [Fe/H]=$-0.6 \pm 0.2$.
The isochrones fit the color and shape of the turnoff
and subgiant branch so precisely that we can constrain
the metallicity as well as the age.
The same isochrones
also fit SL 663 and NGC 2155, although our photometry for
these clusters has much larger errors.  We find these
clusters to be 0.8 Gyr younger, and 0.4 dex more metal
rich, than recently reported in the literature.
Consequently, we argue that NGC 2121, NGC 2155, and SL 663
are not properly assigned to the age gap in the
LMC, but instead are among the first clusters to
have formed in the relatively metal rich, younger group of
LMC clusters.  We propose a new definition of the LMC Age Gap as
extending from 3.2 to 13 Gyr, with ESO121-SC03 still the only remaining
candidate for membership in the age gap.

\end{abstract}


\section{Introduction}

In the course of reducing data from our snapshot survey of
Magellanic globular clusters (GO-5475; PI Shara),
we noticed three clusters in the
LMC with peculiar color-magnitude diagrams:  they appeared to
have two turnoff points, complete with subgiant branches.
Without improved photometry, it was not possible (in our
opinion) to determine an age for these clusters.  The
doubled turnoffs and subgiant branches appeared so evident to us
that exotic explanations (cluster mergers, multiple bursts
of star formation with in a cluster) would have to be considered,
if the effect were real.
The clusters might be 1-3 Gyr old metal rich clusters, or they
could be older than 4 Gyr, perhaps lying in the
in the  4-12 Gyr ``age gap''
of LMC clusters (Jensen, Mould, \& Reid 1988; Da Costa 1991; van den Bergh 1991).

After our LMC snapshot data became public, Sarajedini (1998)
argued that NGC 2121, 2155 and SL6633 are
old LMC clusters with [Fe/H]$\approx -1$ that properly should
be assigned to the age gap.  Motivated by the peculiar results
of our own reductions,
we sought
to obtain much longer
integrations of these clusters using HST, and
the TAC granted observations of one target, NGC 2121.

The formation history of clusters in the LMC is known to be sporadic.  Jensen,
Mould, \& Reid (1988) were unable to find any LMC clusters (other than
ESO 121-SC03) with ages between 4 and 10 Gyr; they proposed the existence
of a gap in the cluster age distribution.
In considering the ages and metallicities of LMC clusters, Olszewski et al.
(1991)
show that a  gap is present in both age and metallicity between those younger
clusters with ages in the range 1-3 Gyr, and very old globular clusters similar
to those found in the Milky Way.  The recent photometry of Olsen et al. (1998)
and Johnson et al. (1999) strengthens further the existence of the age and
metallicity gap:
the oldest LMC clusters are indeed excellent matches for old Milky Way halo globular
clusters such
as M3 and M5. The younger LMC clusters have [Fe/H]$<-1$; it is interesting that
ground based photometry of NGC 1754
resulted in an erroneous (young) age for this
metal poor cluster; Olszewski et al. suspected that it might be old, and new HST
photometry now clearly places it in the very old group of clusters.  Although
less prominently discussed in the literature, the metallicity gap is just as
evident as the age gap.  While clusters of intermediate metallicity $\sim -1.4 $
dex are known in the SMC (Da Costa \& Hatzidimitriou 1988)
they have never been found
in the LMC, even using the modern Ca triplet method employed by Olszewski et al.
(1991). No further results have been found to challenge the gap in metallicity
between the younger group of clusters at $\sim 0.7\pm 0.3$ dex, and the oldest
LMC globular clusters at $-2$ dex. Only ESO121-SC03 remains a strong candidate
for a cluster lying firmly in the age/metallicity gap, at 10 Gyr old (Mateo
et al. 1986) and
[Fe/H]=$-0.91\pm 0.16$ from high resolution spectroscopy (Hill et al. 2000).
The reality of the age gap has been most securely established by
recent photometric surveys of large numbers of clusters
(e.g. Geisler et al. 1997, Bica et al. 1998; Geisler et al. 1999).
These studies find virtually no new candidate clusters that might
lie in the gap.

\subsection{Observations}

We imaged NGC 2121 on 28 January 2000 using WFPC2 for 1600 sec in each of
F555W and F814W, and for 800 sec in F336W. The frames were reduced using the
standard pipeline procedures. Photometry was obtained using DAOPHOT/ALLFRAME
(Stetson 1994) and was calibrated using the new transformation equations and
corrections of Dolphin (2000). These transformations and corrections account
for the charge transfer efficiency and the pre/post-cool down differences.
Application of the Dolphin corrections accounts for minor differences
between our photometry of the snapshot clusters,
and that of Sarajedini (1998).  For NGC 2121, of course, our improved
color-magnitude diagram is due to our longer integration time.

Several clusters of interest (NGC 2121, 2155, 2193, SL 556, and SL 663) were
observed in 1994, as part of our snapshot survey (PI M. Shara).  The exposure
times and dates of observations for these clusters are given in Table 1.
All of the snapshot data were reduced as described above, but because the
snapshot data have only one frame in each color, the signal-to-noise is smaller
and the cosmic ray and hot pixel contamination is larger for those data.

\section{Color-magnitude Diagrams}

Figure \ref{n2121visub} shows the color-magnitude
diagram of NGC 2121, based on our new data.  We illustrate
statistically subtracted data, as well as the CMD formed
from the ``field'' population constructed from the outer
portions of the WF chips (half the total area of WFPC2). 
Though the cluster is likely still present in this ``field'' population
it serves as a clear upper limit to the field contamination. 
The true nature of the apparent doubled turnoff is now obvious: there is a
wide hydrogen core exhaustion gap at the main sequence turnoff
point.  This gap occurs when the completely convective core
suddenly exhausts hydrogen, requiring a structural readjustment
of the star which has the effect of causing the blueward hook
in the evolutionary tracks. The strong descending subgiant
branch (due to the high metallicity) caused some confusion in
the lower S/N data.  The turnoff morphology we observe is typical 
of Galactic open clusters of approximately this age (e.g. M67; 
Montgomery et al. 1993). Rosvick \& VandenBerg (1998), show that
Galactic open clusters with a similar redward curvature of the main sequence
turnoff just below the subgiant branch can be fit using
models with $\approx 50\%$ of the convective overshooting predicted by
the Roxburgh (1978) criterion.
A detailed fit of our data to such models will be deferred to
a later paper; in this paper we focus on the
question of whether NGC 2121 and similar
clusters actually fall well within the age/metallicity gap.

We achieve an excellent fit of the Girardi et al. (2000) isochrones 
to the data
(Figure \ref{n2121vi}), for an age
of 3.2 Gyr and [Fe/H]=$-0.68$.  These isochrones are based on
models with a moderate amount of convective overshooting.
We make the fit by tying the isochrone red
clump to the observed red clump (as in Rich et al. 2000), which removes
uncertainties such as the spatial depth of the LMC cluster system.  
The goodness of
the fit, including the accurate reproduction of the subgiant branch, is
remarkable.  Figure \ref{n2121uv} shows that the fit is equally good using
our new $U,V$ photometry.  Figure \ref{n2121vi}
proves that the fit strongly depends on metallicity, and while Olszewski et al's (1991) spectroscopy gives
$-0.6\pm 0.1$ dex, the isochrone fit also requires this metallicity.

The excellent isochrone fit to the new color-magnitude diagram
of NGC 2121 gives a foundation on which to approach the
snapshot data, which have much larger photometric errors.
We find that the same isochrones $(\log t=9.5,
\rm [Fe/H]=-0.68) $ that fit NGC 2121 also
appear to fit NGC 2155 and SL 663 as well.
Figure \ref{n2121group} shows isochrones overlaid on the snapshot data,
fitting at the red clump.
As suggested by Sarajedini (1998), NGC 2121, 2155, and SL 663
may be very similar in age.  Nearly coeval intermediate
age star clusters are also known in the SMC (Rich et al. 2000).
We also apply our isochrone fits to NGC 2193 and SL 556 (figure \ref{n21group2})
and again find ages of approximately 2 Gyr (Table 2).
Having now understood NGC 2121 as a single population of intermediate
age, we have more confidence our isochrone fit for the other clusters,
and we are confident that none are older than NGC 2121.

\section{Discussion}

Do NGC 2121, NGC 2155, and SL 663 belong in the age gap, as suggested by
Sarajedini (1998), or are they instead the oldest clusters in the
younger group of LMC clusters?  Figures \ref{clustlinage} and \ref{clustlogage}
illustrate the
age-metallicity relationship for LMC clusters.   These plots show
additional data points for 
the young clusters (Dirsch et al. 2000) which are not part of this
study.  We also illustrate the old Magellanic clusters as being coeval
with the Milky Way halo globular clusters
(Olsen et al. 1998; Johnson et al. 1999).   

The gap in age and metallicity remains apparent.
The oldest LMC clusters are all very metal poor, approximately
[Fe/H]=$-2$, as is the case for comparable old clusters in the
Milky Way.  In contrast, only a few clusters in the younger group
have $\rm [Fe/H]<-1$.
Although it may be a matter of semantics
more than astrophysics,  the combination of our slightly
younger ages and our choice to adopt the Olszewski et al. spectroscopic
metallicity for NGC 2121, NGC 2155, and SL 663 unambiguously places them
out of the age gap and associates them with the younger clusters.
It is interesting that the LMC interstellar medium was apparently enriched from
[Fe/H]$\sim -2$ to [Fe/H]$\sim -0.5$ without leaving behind long
lived star clusters.  It is clear that the formation of star clusters
may accompany chemical enrichment, but that it is not a requirement
for enrichment to occur.

Can we defend our claim that the metallicity of NGC 2121 is
high based on the isochrones alone?  All published spectroscopic
metallicity measurements in NGC 2121 give metallicities higher
than $-1$ dex.
Cohen (1982)
derives [Fe/H]=$-0.95$ from Fe, Ca, Na, and Mg line widths at
low resolution, for two stars.  Bica et al. find [Fe/H]=$-0.75$
from integrated DDO photometry of the cluster, but such a method
can be affected by field contamination, a concern also for
the integrated light spectral indices of de Freitas Pacheco  et al.
1998).   We consider the measurement of [Fe/H]=$-0.61$ by Olszewski et al.
(based on Ca triplet spectra of two stars) to be the most reliable,
because the Ca triplet method is well calibrated, and the composition
is scaled Solar for metal rich stars.  Furthermore,
Olszewski et al. confirm radial velocity membership of the
two stars in  NGC 2121 and use calibrating clusters with higher
metallicity.   Olszewski et al. also measure the Ca triplet
in NGC 2155 finding [Fe/H]=$-0.55$.  It is interesting to note that
Olszewski et al. find [Fe/H]=$-0.93$ for ESO 121-SC03 using the low
resolution Ca triplet method, while Hill et al. 2000 find [Fe/H]=$-0.91$
using high resolution VLT spectroscopy of stars in the cluster.  We believe
that the Ca triplet metallicities are more accurate than abundance
measurements derived from fits to the red giant branch slope.

The isochrone fits at the turnoff (clear descending
subgiant branch) also confirm the high metallicity
of NGC 2121 and the other clusters in this group.
For these clusters,
we now assign [Fe/H]=$-0.6$, 0.4 dex higher than
derived from the red giant branch slope by Sarajedini (1998). The
descending subgiant branch is also seen in intermediate
age, relatively metal rich Galactic clusters, and is caused
by an increase in blanketing, when the star's atmosphere expands and cools
in the approach to the red giant branch.  The concurrence of abundance
inferred from the color-magnitude diagrams and the Ca triplet spectroscopy
compels us to favor the higher abundance
scale for NGC 2121, 2155, and SL663.  We are convinced
that these clusters are more metal rich than ESO 121-SCO3, and by
metallicity as well as age, belong on the young side of the age gap.

The core of our argument associating these clusters with the younger
group (rather than the age gap) rests on both a +0.4 dex increase
in metallicity and a 0.2 dex decrease in age relative to Sarajedini's
(1998) values. However, our new
values are supported by the data and, and a combination of
$\log t=9.5$ and [Fe/H]=$-1$ simply does not fit the color-magnitude
diagram of NGC 2121.  Plotted in linear space, the gap between our
age of 3.2 Gyr and the 13 Gyr ages for the oldest LMC clusters
(or even the 10 Gyr age of ESO 121-SCO3) is still very large.
We propose the former -- the interval from 3.2 Gyr to 13 Gyr -- as the
new boundaries for the LMC age gap, verified by high precision WFPC2
photometry.
We conclude that the age gap in the LMC remains real, and unexplained.

\begin{acknowledgements}

Support for this work was provided by NASA through grant GO-8141 from the
Space Telescope Science Institute, which is operated by AURA, Inc., under
NASA contract NAS 5-2655. R.M.R. appreciates the hospitality of the Hayden
Planetarium at the American Museum of Natural History.  We thank
Ivo Saviane, Alvio Renzini, Gary Da Costa, and
Don VandenBerg for valuable comments.

\end{acknowledgements}

\begin{figure}

\caption[junk]{\label{n2121visub} The globular cluster NGC 2121
fills the entire field of WFPC2.  We have isolated a ``field''
population from a field covering half the area of the WFPC2,
at the greatest distance from the cluster center in the
PC chip (Right panel).  This ``field'' color-magnitude diagram
has been statistically subtracted from the complete data set, to yield
our best effort at representing the true cluster color-magnitude diagram (Left Panel). }

\end{figure}

\begin{figure}

\caption[junk]{\label{n2121vi} Left panel: Color-magnitude diagram of the
cluster NGC 2121 from new WFPC2 imagery; PC data only with F555W and F814W
data transformed to Johnson $V,I$. The redward arc of the turnoff point, with
the blue hook leading to the subgiant branch are indicative of
hydrogen core exhaustion in intermediate-mass
stars.  The descending subgiant branch is indicative of
high metallicity.  We overlay the Girardi et al. (2000) isochrones,
forcing the fit to the red clump stars.  Right panel: NGC 2121, complete
data (including the field main sequence).  We overlay the $\log t=9.5$
isochrones, but varying the metallicity. Notice that [Fe/H]=$-0.68$ 
is a superior fit to the data.}

\end{figure}

\begin{figure}
\caption[junk]{\label{n2121uv} Color-magnitude diagram of the cluster NGC 2121
using new F336W and F555W imagery, transformed to Johnson $U,V$. Girardi et
al. (2000) isochrones are overlaid on the data.}
\end{figure}

\begin{figure}
\caption[junk]{\label{n2121group}  Color-magnitude diagrams for NGC 2121,
2155, and SL663 from the short snapshot survey exposures.  These data are
typically 120-230 sec exposures with no cosmic ray cleaning.  The best fit
isochrones $(\log t=9.5, \rm [Fe/H]=-0.68)$ from the high quality CMD of
NGC 2121 are overlaid on these much noisier CMDs.  We conclude that there
is no evidence for any clusters in this group being older than NGC 2121.}
\end{figure}

\begin{figure}
\caption[junk]{\label{n21group2} Color-magnitude diagrams for two additional
candidate old LMC clusters in our sample, NGC 2193 and SL 556, also fit using
the best-fit Girardi et al. (2000) isochrones ($\log t=9.3$ to 9.7,
in steps of 0.1) with [Fe/H]=$-0.68$.  We find an age of $2.2+/-0.5$ Gyr
for these clusters; Table 2.  These
clusters are likely to be younger than NGC 2121.}
\end{figure}

\begin{figure}
\caption[junk]{\label{clustlinage} Plot of age as a function of metallicity
for LMC clusters, based on the data in Dirsch et al. (2000).  Our new data
(given in Table 2) 
are indicated by a cross (our best fit to NGC 2121, which we
apply to NGC 2155 and SL663) and the plus (our ages for NGC 2193 and SL556).
Ages and metallicities for the clusters are given in Table 2.  
We estimate an error of $\pm 0.05$ in $\log$ age from our fit of the
isochrones.  
We adopt the metallicity scale of Olszewski et al.
(1991) for our clusters (see text).  
We also adopt the metallicity of Hill et al. (2000)
for ESO121-SC03, and we presume an age of 13 Gyr for the oldest group of LMC
globular clusters, based on recent WFPC2 photometry.  Notice that the
age gap extends from 3.3 to 13 Gyr.}
\end{figure}

\begin{figure}
\caption[junk]{\label{clustlogage} Plot of age as a function of metallicity
for LMC clusters as in Figure 5, 
based on the data in Dirsch et al. (2000) but using a
logarithmic age scale. We have modified the Dirsch et al. (2000) data as
discussed in Figure \ref{clustlinage}.   Our new data
(given in Table 2) 
are indicated by a cross (our best fit to NGC 2121, which we
apply to NGC 2155 and SL663) and the plus (our ages for NGC 2193 and SL556).
Ages and metallicities for the clusters are given in Table 2.  The gap remains very clear, with the WFPC2 data
confirming the findings of recent ground-based surveys that the age gap is
real.}
\end{figure}

\begin{table}[h] 
\begin{centering}
\vbox{\caption{HST/WFPC2 observations used in this study}
\label{jobs}
\vspace{\baselineskip}
\begin{tabular}{lcccc}
\hline\hline
\vspace{3pt}
\strut  Cluster & Filter & Total Exposure time & \# of Exposures & Date \\
 & & (Seconds) & & (dd/mm/yyyy) \\
\strut \\ \hline
& & & & \\
NGC 2121 & F450W & 230.0 & 1 & 02/02/1994 \\
         & F555W & 120.0 & 1 & 02/02/1994 \\
         & F555W & 1600.0 & 4 & 28/01/2000 \\
         & F814W & 1600.0 & 4 & 28/01/2000 \\
         & F336W & 800.0 & 2 & 28/01/2000 \\
& & & & \\
NGC 2193 & F450W & 230.0 & 1 & 30/01/1994 \\
         & F555W & 120.0 & 1 & 30/01/1994 \\
& & & & \\
NGC 2155 & F450W & 230.0 & 1 & 01/02/1994 \\
         & F555W & 120.0 & 1 & 01/02/1994 \\
& & & & \\
SL 556   & F450W & 230.0 & 1 & 01/02/1994 \\
         & F555W & 120.0 & 1 & 01/02/1994 \\
& & & & \\
SL 633   & F450W & 230.0 & 1 & 01/02/1994 \\
         & F555W & 120.0 & 1 & 01/02/1994 \\

\strut \\ \hline
\end{tabular}
}
\end{centering}
\end{table}

\begin{table}[h] 
\begin{centering}
\vbox{\caption{Cluster Ages and [Fe/H] from this study}
\label{clage}
\vspace{\baselineskip}
\begin{tabular}{lcccc}
\hline\hline
\vspace{3pt}
\strut  Cluster & [Fe/H] & Age & $\log$ Age \\
 & & (Gyr) & $\log (yr)$ \\
\strut \\ \hline
& & & \\
NGC 2121 & -0.68 & 3.2 & 9.5 \\
NGC 2155 & -0.68 & 3.2 & 9.5 \\
NGC 2193 & -0.68 & 2.2 & 9.35 \\
SL 556 & -0.68 & 2.2 & 9.35 \\
SL 633 & -0.68 & 3.2 & 9.5 \\
\strut \\ \hline
\end{tabular}
}
\end{centering}
\end{table}

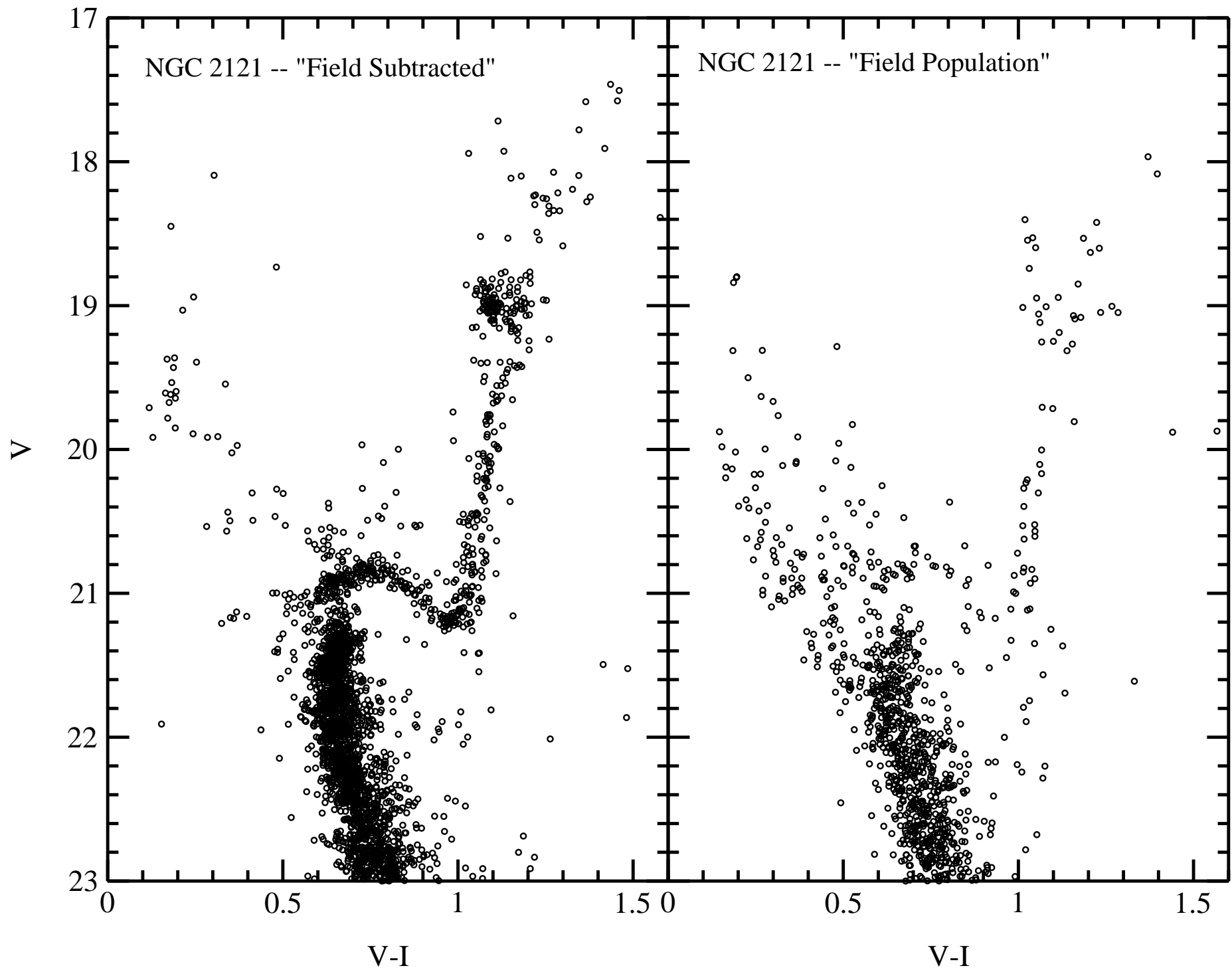
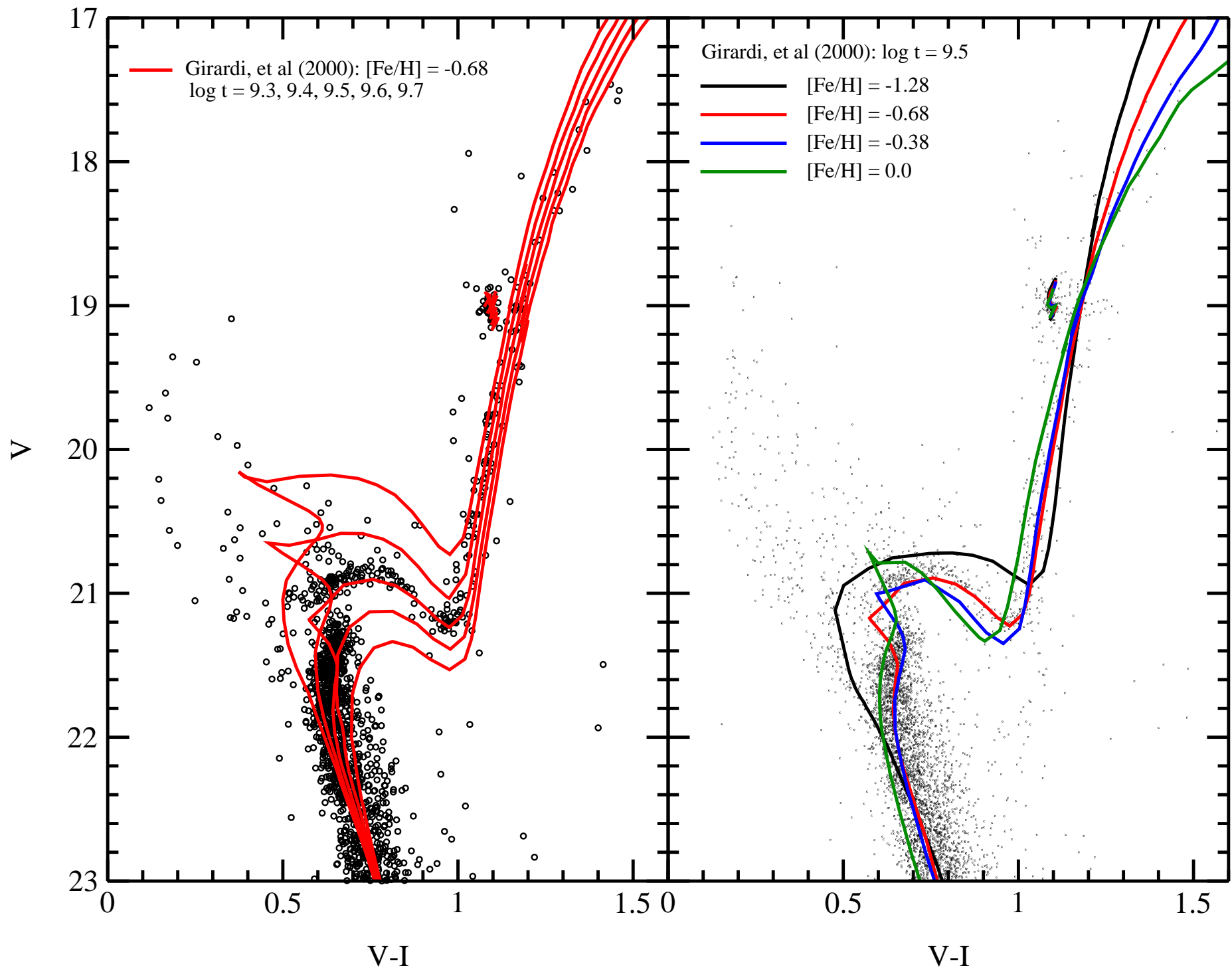
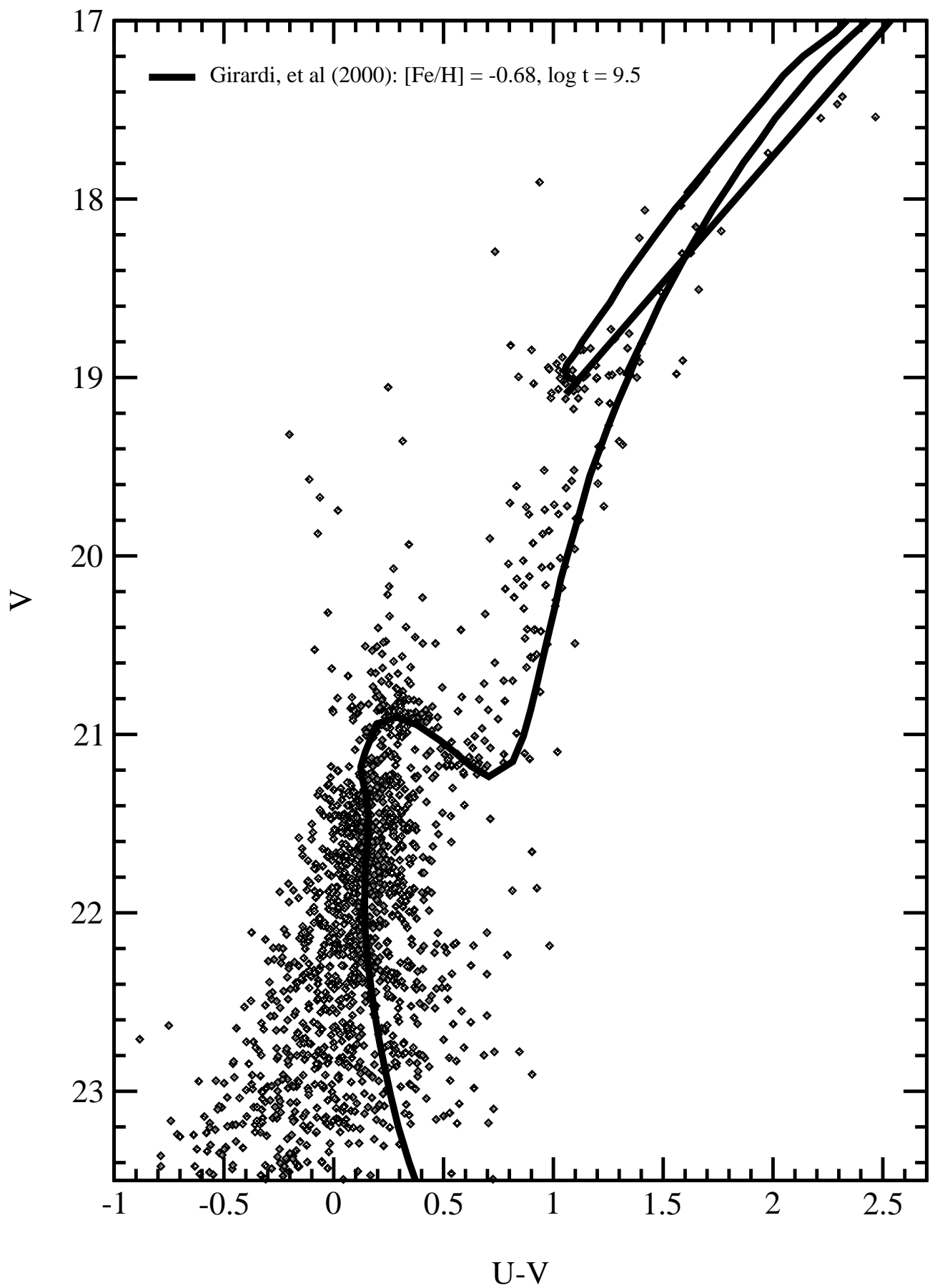
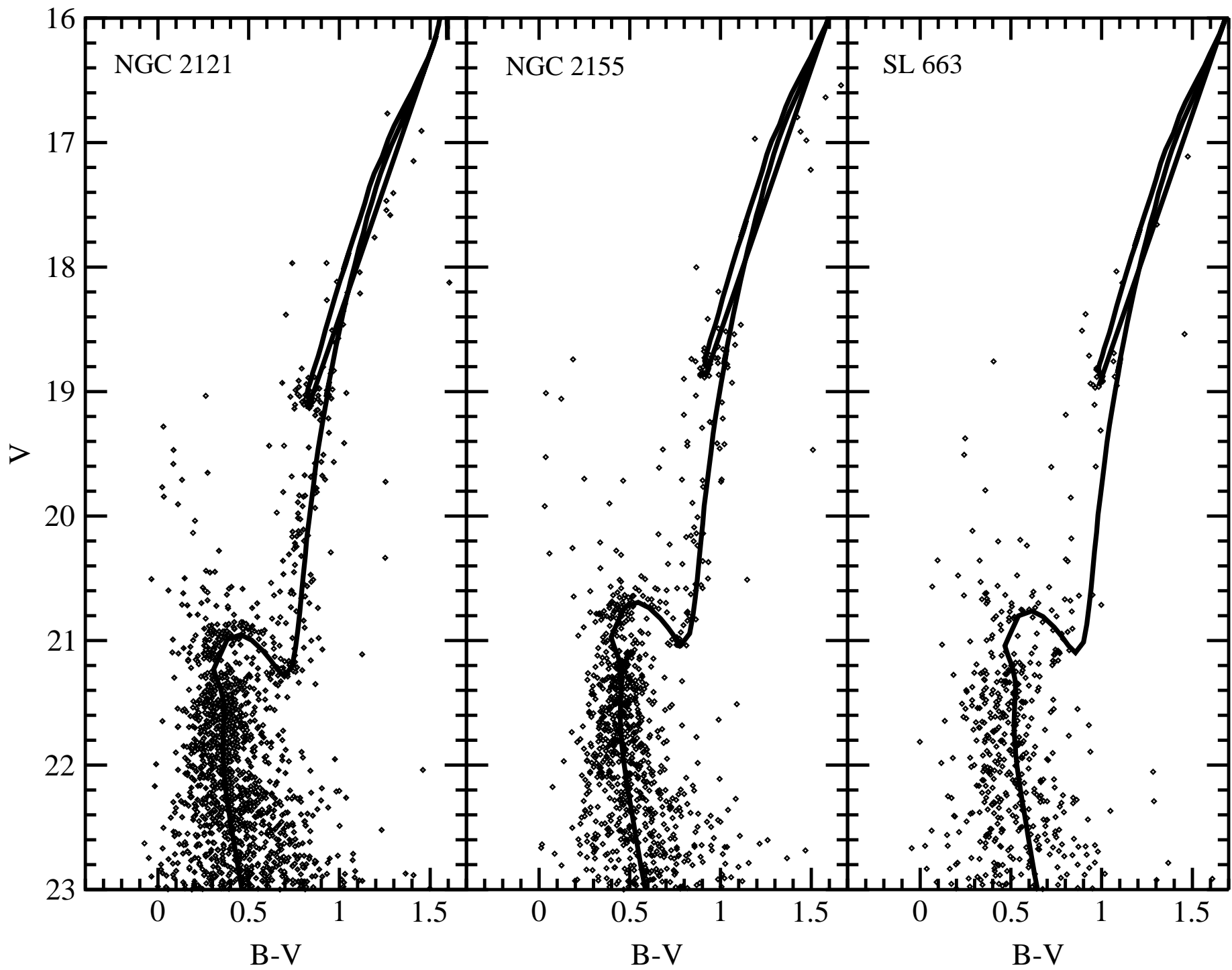
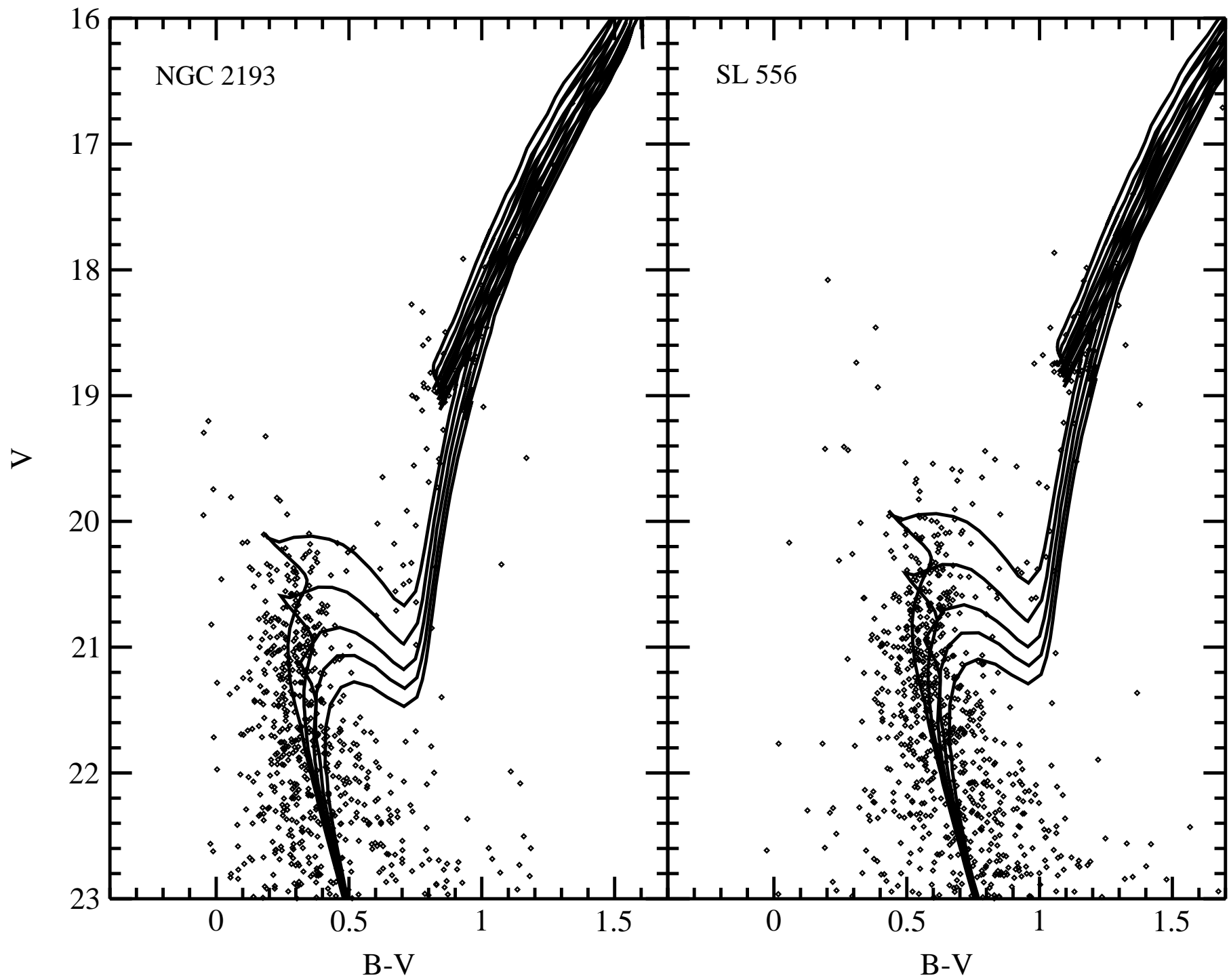
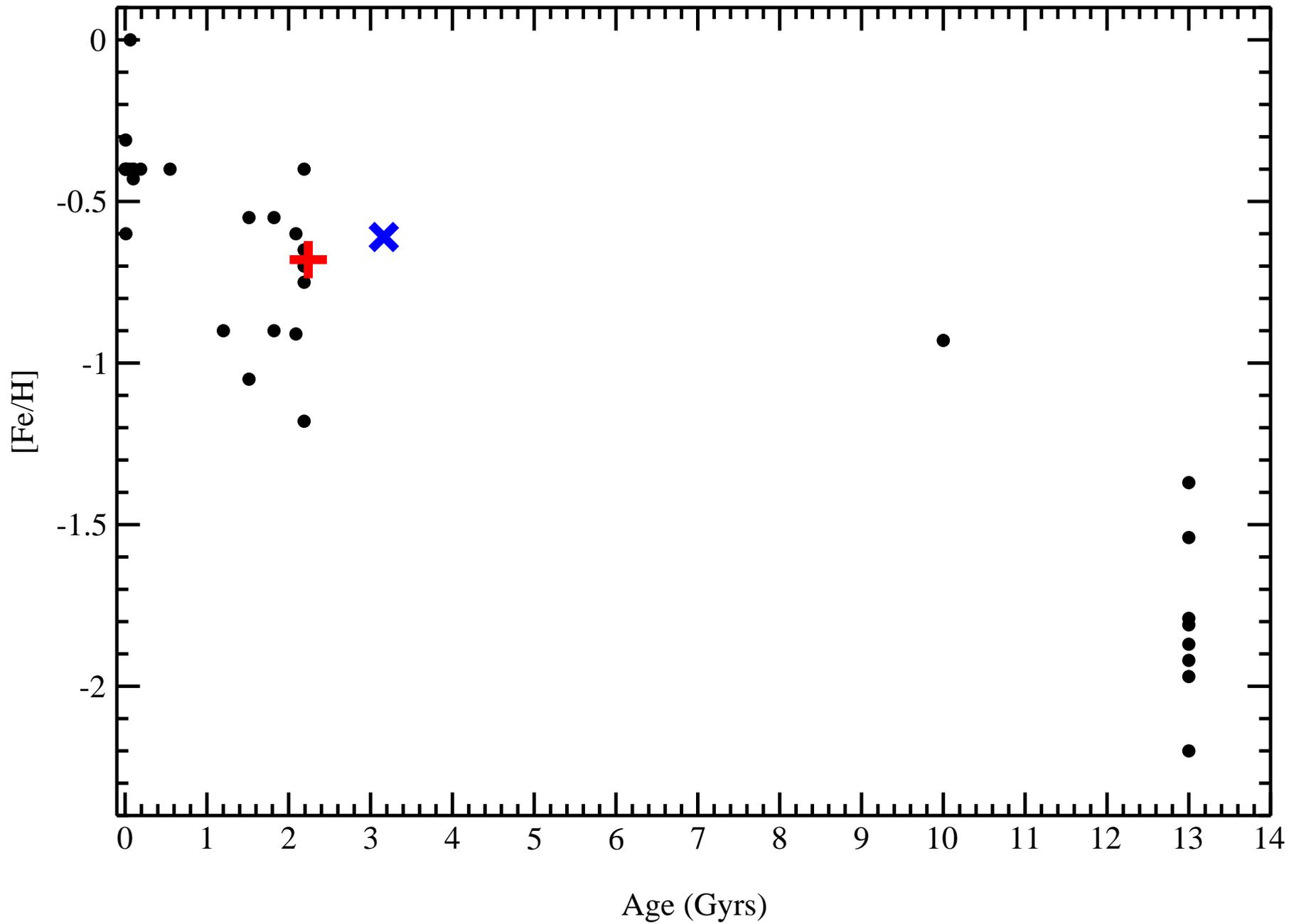
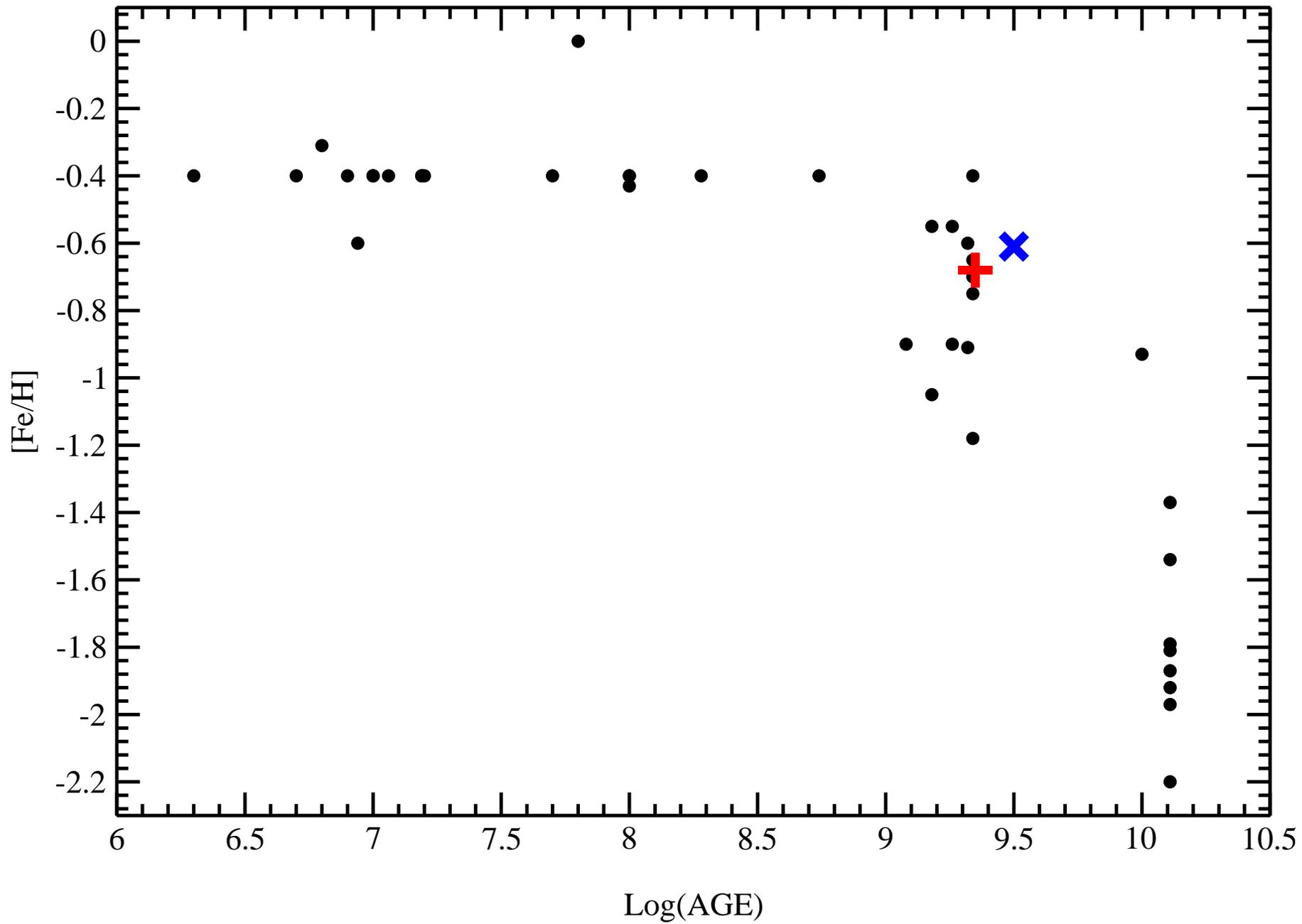
\end{document}